\documentclass[10pt,conference]{IEEEtran}  
\usepackage{amsmath,amsthm, epsfig, verbatim, color, amsfonts} 
\usepackage{graphicx}
\title{Feedback Communication over Individual Channels}
\author{Yuval Lomnitz, Meir Feder \\
Tel Aviv University, Dept. of EE-Systems \\
Email: \{yuvall,meir\}@eng.tau.ac.il}

\theoremstyle{plain}
\newtheorem{theorem}{Theorem}

\theoremstyle{definition}

\def\vr{\mathbf}

\def\Pr{\mathrm{Pr}}
\def\Normal{\mathcal{N}}

\def\Remp{R_{\mathrm{emp}}}

\def\half{\frac{1}{2}}

\newcommand{\arrowexpl}[1] {\raisebox{-1.0ex}{$\stackrel{\textstyle \longrightarrow}{\scriptscriptstyle #1}$}}

\def\RLBONE{R_{\mathrm{LB1}}}
\def\RLBTWO{R_{\mathrm{LB2}}}

{\hspace{\stretch{1}} \textcolor{red}{$\clubsuit^{-1}$ }}

\begin{document}
\maketitle

\begin{abstract} 
We consider the problem of communicating over a channel for which no mathematical model is specified. We present achievable rates as a function of the channel input and output sequences known a-posteriori for discrete and continuous channels. Furthermore we present a rate-adaptive scheme employing feedback which achieves these rates asymptotically without prior knowledge of the channel behavior.
\end{abstract}


\IEEEpeerreviewmaketitle

\section{Introduction}
The problem of communicating over a channel with an individual, predetermined noise sequence which is not known to the sender and receiver was addressed by Shayevitz and Feder \cite{Ofer_BSC}\cite{Ofer_EMP} and Eswaran et. al. \cite{Eswaran}\cite{Eswaran_conf}. The simple example discussed in \cite{Ofer_BSC} is of a binary channel $y_n=x_n \oplus e_n$ where the error sequence $e_n$ can be any unknown sequence. Using perfect feedback and common randomness, communication is shown to be possible in a rate approaching the capacity of the binary symmetric channel (BSC) whose the error probability equals the empirical error probability of the sequence (the relative number of '1'-s in $e_n$). Subsequently both authors extended this model to general discrete channels and modulu-additive channels (\cite{Eswaran},\cite{Ofer_EMP} resp.) with an individual state sequence, and showed that the empirical mutual information can be attained.

In this work we take this model one step further. We consider a channel where no specific probabilistic or mathematical relation between the input and the output is assumed. We term this channel an \textit{individual channel} and we would like to characterize the achievable rate using only the input and output sequences. The decoder may have a feedback link in which the channel output or other information from the decoder can be sent back. Without this feedback it would not be possible to match the rate of transmission to the quality of the channel so outage would be inevitable. This model has various advantages and disadvantages compared to the classical one, however there is no question about the reality of the model: this is the only channel model that we know for sure exists in nature. This point of view is similar to the approach used in universal source coding of individual sequences where the goal is to asymptotically attain for each sequence the same coding rate achieved by the best encoder from a model class, tuned to the sequence.

Just to inspire thought, let us ask the following question: suppose the sequence $\{x_i\}_{i=1}^n$ with power
$P = \frac{1}{n}\sum_{i=1}^{n}{x_i^2}$ encodes a message and is transmitted over a continuous real-valued input channel. The output sequence is $\{y_i\}_{i=1}^n$. One can think of $v_i = y_i - x_i$ as a noise sequence and measure its power $N=\frac{1}{n}\sum_{i=1}^{n}{v_i^2}$. The rate $R=\half \log \left( 1 + \frac{P}{N} \right)$ is the capacity of a Gaussian additive channel with the same noise variance. Is the rate $R$ also achievable in the individual case, under appropriate definitions ?

The way it was posed, the answer to this question would be "no", since this model predicts rate of $\half$ bit/use for the channel whose output $\forall i: y_i=0$ which cannot convey any information. However with the slight restatement done in the next section (see Eq.(\ref{R_emp_continuous}) below) the answer would be "yes".

We consider two classes of individual channels: discrete input and output channels and continuous real valued input and output channels. In both cases we assume that feedback and common randomness exist (perfect feedback is not required). In \cite{FullPaper} we address also the case where feedback does not exist, which yields interesting results, but to keep the presentation concise we focus here on the more important case of feedback communication. The main result is that with small amount of feedback, a communication at a rate close to the empirical mutual information (or its Gaussian equivalent for continuous channels) can be achieved, without any prior knowledge, or assumptions, about the channel structure. Here we present the main result and the communication scheme obtaining it and give an outline of the proof. The full proof is omitted and appears in \cite{FullPaper}. We also give several examples and highlight areas for further study.


\section{Overview of the main results obtained so far}\label{sec:overview}
We start with a high level overview of the definitions and results. The discussion below is conceptual rather than accurate, while the detailed definitions follow in the next section.

We say a given rate function $\Remp : \mathcal{X}^n \times \mathcal{Y}^n \to \mathbb{R}$ is achieved by a communication scheme with feedback if for large block size $n$, data at rate close to or exceeding $\Remp(\vr{x}, \vr{y})$ is decoded successfully with arbitrarily large probability for every output sequence and almost every input sequence. Roughly speaking, this means that in any instance of the system operation, where a specific $\vr x$ was the input and a specific $\vr y$ was the output, the communication rate had been at least $\Remp(\vr x, \vr y)$. Note that the only statistical assumptions are related to the common randomness, and we consider the rate (message size) and error probability \textit{conditioned} on a specific input and output, where the error probability is averaged over common randomness.

The definition of achievability is not complete without stating the input distribution, since it affects the empirical rate. For example, by setting $\vr x = 0$ one can attain every rate function where $\Remp(0,\vr y)=0$ in a void way, since other $\vr x$ sequences will never appear. In contrast with classical results of information theory, we do not use the input distribution only as a means to show the existence of good codes: taking advantage of the common randomness we require the encoder to emit input symbols that are random and distributed according to a defined prior (currently we assume i.i.d. distribution).

In this paper we focus on rate functions that depend on the instantaneous (zero order) empirical statistics. Extension to higher order models seems technical. For the discrete channel we show that a rate
\begin{equation}\label{R_emp_discrete}
\Remp = \hat{I} (\vr{x}; \vr{y})
\end{equation}
is achievable with any input distribution $Q(x)$ where $\hat{I}(\cdot;\cdot)$ denotes the empirical mutual information \cite{Goppa}. For the continuous (real valued) channel we show that a rate
\begin{equation}\label{R_emp_continuous}
\Remp = \half \log \left( \frac{1}{1-\hat\rho(\vr x, \vr y)^2} \right)
\end{equation}
is achievable with Gaussian input distribution $\Normal(0,P)$, where $\hat\rho \equiv \frac{\vr x^T \vr y}{\lVert \vr x \rVert \lVert \vr y \rVert }$ is the empirical correlation factor between the input and output sequences (at this stage for simplicity $\hat\rho$ is defined in a slightly non standard way without subtracting the mean). Although the result regarding the continuous case is less tight, we show in \cite{FullPaper} that this is the best rate function that can be defined by second order moments, and it is tight for the Gaussian additive channel (for this channel $\rho^2 = \frac{P}{P+N}$ therefore $\Remp = \half \log \left( 1 + \frac{P}{N}\right)$). The same rates apply also to the case of communication without feedback where achievability is defined by the ability to decode a fixed rate $R$ whenever $\Remp > R$.

We may now rephrase our example question from the introduction so that it will have an affirmative answer: given the input and output sequences, describe the output by the virtual additive channel with a gain $y_i = \alpha x_i + v_i$, so the effective noise sequence is $v_i = y_i - \alpha x_i$. Chose $\alpha$ so that $\vr v \perp \vr x$, i.e. $\frac{1}{n} \sum_i{v_i x_i}=0$. An equivalent condition is that $\alpha$ minimizes $\lVert \vr v \rVert^2$. The resulting $\alpha$ is the LMMSE coefficient in estimation of $\vr y$ from $\vr x$ (assuming zero mean), i.e. $\alpha = \frac{\vr x ^T \vr y}{\lVert \vr x \rVert^2}$. Define the effective noise power as $N=\frac{1}{n}\sum_{i=1}^{n}{v_i^2}$, and the effective $\textit{SNR} \equiv \frac{\alpha^2 P}{N}$. It is easy to check that $\textit{SNR}=\frac{\hat\rho^2}{1-\hat\rho^2}$. Then according to Eq.(\ref{R_emp_continuous}) the rate $R=\half \log \left( 1 + \textit{SNR} \right)$ is achievable, in the sense defined above. Reexamining the counter example we gave above, in this model if we set $\vr y = 0$ we obtain $\hat\rho=0$ and therefore $\Remp=0$, or equivalently the effective channel has $\vr v = 0$ and $\alpha=0$, therefore $\textit{SNR}=0$ (instead of $\vr v = -x$, $\alpha=1$ and $\textit{SNR}=1$).

As will be seen, we achieve these rates by random coding and universal decoders, and use iterated instances of rateless coding. The scheme is able to operate asymptotically with "zero rate" feedback (meaning any positive capacity of the feedback channel suffices). A similar although more complicated scheme was used in \cite{Eswaran}. The main differences are the use of training to evaluate the stopping condition as well as a different code construction and are summarized in \cite{FullPaper}.

The classical point of view first assumes a channel model and then devises a communication system optimized for it. Here we take the inverse direction: we devise a communication system without assumptions on the channel which guarantees rates depending on channel behavior. The channel model we assume is more stringent than the probabilistic and semi-probabilistic models since we make less assumptions about the channel, and the error probability and rate are required to be met for (almost) every input and output sequence (rather than on average). This change of viewpoint does not make probabilistic or semi probabilistic channel models redundant but merely suggests an alternative. By using a channel model we can formalize questions relating to optimality such as capacity (single user, networks) and error exponent as well as guarantee a communication rate a-priori. Another aspect is that we pay a price for universality. Even if one considers an individual channel scheme that guarantees asymptotically optimum rates over a large class of channels, it can never consider all possible channels (block-wise), and for a finite block size it will have larger overhead (a reduction in the amount of information communicated with same error probability) compared to a scheme optimized for the specific channel.

Several concepts used in this work such as common randomness and rateless coding, are borrowed from prior work on arbitrarily varying channels (AVC, see for example \cite{Lapidoth_AVC}\cite{Csiszar_AVC}) compound channels with feedback \cite{Shulman}\cite{Tchamkerten} and individual noise sequence channels with feedback \cite{Ofer_EMP}\cite{Eswaran}. It is worth noting \cite{Agarwal_RD} where a somewhat similar concept was used in defining an achievable communication rate by properties of the channel input and output. An important observation is that a strict definition of capacity exists only for fixed rate systems (where the capacity is the supremum of achievable rates) while in rate adaptive communication there is some freedom in determining the rate function.

Following our results, the individual channel approach becomes a very natural starting point for determining achievable rates for various probabilistic and semi-probabilistic models (AVC, individual noise sequences, probabilistic models, compound channels) under the realm of randomized encoders, since the achievable rates for these models follow easily from the achievable rates for specific sequences, and the law of large numbers. We will give some examples later on.

\section{Definition of variable rate communication system with feedback}\label{sec:definitions}
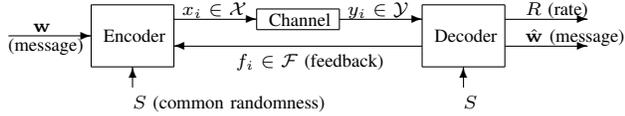
\begin{figure}[t]
\setlength{\unitlength}{0.55mm}
\scriptsize
\hspace{\stretch{1}}
\begin{picture}(140, 30)
\put(23,16){Encoder}\put(20,10){\line(1,0){20}}\put(40,10){\line(0,1){15}}\put(40,25){\line(-1,0){20}}\put(20,25){\line(0,-1){15}}
\put(63,20){Channel}\put(60,19){\line(1,0){20}}\put(80,19){\line(0,1){5}}\put(80,24){\line(-1,0){20}}\put(60,24){\line(0,-1){5}}
\put(103,16){Decoder}\put(100,10){\line(1,0){20}}\put(120,10){\line(0,1){15}}\put(120,25){\line(-1,0){20}}\put(100,25){\line(0,-1){15}}
\put(0,17.5){\vector(1,0){20}}\put(6,18.5){$\vr w$}\put(0,14){(message)}
\put(40,21.5){\vector(1,0){20}}\put(42,22.5){$x_i \in \mathcal{X}$}
\put(80,21.5){\vector(1,0){20}}\put(82,22.5){$y_i \in \mathcal{Y}$}
\put(100,15){\vector(-1,0){60}}\put(55,10){$f_i \in \mathcal{F}$ (feedback)}
\put(120,21.5){\vector(1,0){20}}\put(125,22.5){$R$ (rate)}
\put(120,15){\vector(1,0){20}}\put(125,16){$\hat{\vr w}$ (message)}
\put(30,5){\vector(0,1){5}}\put(30,0){$S$ (common randomness)}
\put(110,5){\vector(0,1){5}}\put(110,0){$S$}
\end{picture}
\hspace{\stretch{1}}
\caption{Rate adaptive encoder-decoder pair with feedback}\label{fig:system_adaptive}
\end{figure}

A randomized block encoder and decoder pair for the channel $\mathcal{X \to Y}$ (defined by the two alphabets $\mathcal{X,Y}$) with block length $n$ adaptive rate and feedback communicates a message expressed by the infinite sequence $\vr w_1^{\infty} \in \{0,1\}^{\infty}$. The system is defined using a random variable $S$ distributed over the set $\mathcal{S}$ (the common randomness) and a feedback alphabet $\mathcal{F}$. The  encoder is defined by a series of mappings $ x_k = \phi_k(\vr w, s, \vr f^{k-1})$ and the decoder is defined by a feedback function $f_k = \varphi_k(\vr y^k, s)$, a decoding function $\hat {\vr w} = \bar\phi (\vr y, s)$ and a rate function $ R = r (\vr y, s)$. The error probability for message $\vr w_1^{\infty}$ is defined as
$P_e^{(\vr w)}(\vr x, \vr y) = \Pr \left({\hat {\vr w}}_1^{\lceil nR \rceil} \neq  {{\vr w}}_1^{\lceil nR \rceil} \big\vert \vr x, \vr y \right)$, i.e. recovery of the first $\lceil nR \rceil$ bits by the decoder is considered a successful reception. This system is illustrated in figure \ref{fig:system_adaptive}.

\section{Statement of the main result}\label{sec:rate_adaptive_theorems}
We consider two cases:
\begin{enumerate}
\item \emph{discrete:} The input and output alphabets $\mathcal{X,Y}$ are discrete and finite, and the prior $Q(x)$ can be arbitrarily chosen
\item \emph{continuous:} The input and output alphabets are real valued $\mathcal{X=Y}=\mathbb{R}$ and the prior is Gaussian $Q = \Normal(0,P)$
\end{enumerate}
The scheme proposed below satisfies the following theorem with respect to these two cases:
\begin{theorem}[Theorems 3,4 of \cite{FullPaper}]\label{theorem:adaptive_rate_combined}
For every $P_e,P_A,\delta,\bar R>0$ there is $n$ large enough and random encoder and decoder with feedback and variable rate over block size $n$ with a subset $J \subset\mathcal{X}^n$, such that:
\begin{itemize}
\item The distribution of the input sequence is $\vr{x} \sim Q^n$ independently of the feedback and message
\item The probability of error is smaller than $P_e$ for any $\vr x, \vr y$
\item For any input sequence $\vr x \not\in J$ and output sequence $\vr{y} \in \mathcal{Y}^n$ the rate is
$R \geq \min \left[ \Remp (\vr{x},\vr{y}) - \delta, \bar R \right]$, where
\begin{equation}\label{eq:def_mu}
\Remp(\vr x, \vr y) \equiv  \left\{ \begin{array}{ll}
\hat I (\vr x, \vr y) & \textrm{discrete}\\
\half \log \left( \frac{1}{1 - \hat\rho^2(\vr x, \vr y)} \right) & \textrm{continuous}\end{array} \right.
\end{equation}
\item The probability of the subset $J$ is bounded by $\Pr(\vr x \in J) \leq P_A$
\end{itemize}
\end{theorem}

The limit $\bar R$, which can be arbitrarily large, reflects the fact the communication rate is finite, even when $\Remp = \infty$ ($\hat\rho^2$ = 1 in the continuous case). In the discrete case $\bar R$ can be omitted (by selecting $\bar R = \log \min(|\mathcal{X}|, |\mathcal{Y}|) \geq \hat I(\vr x, \vr y)$).

Regarding the subset $J$ as we shall see in the proof outline there are some sequences for which poor rate is obtained, and since we committed to an input distribution we cannot avoid them. However there is an important distinction between claiming for example that "for each $\vr y$ the probability of $R < \Remp$ is at most $P_A$" and the claim made
in the theorem that "$R < \Remp$ only when $\vr x$ belongs to a subset $J$ with probability at most $P_A$". The first claim is weaker since choosing $\vr y$ as a function of $\vr x$ may potentially increase the probability of $R < \Remp$ beyond $P_A$, by attempting to select for every $\vr x$ a sequence $\vr y$ for which $\vr x$ is a bad input sequence. This weakness is avoided in the second claim. A consequence of this definition is that the probability of $R < \Remp$ is bounded by $P_A$ for any conditional probability $\Pr(\vr y | \vr x)$ over the sequences. The probability $P_A$ can be absorbed into $P_e$ with the implication that the error probability becomes limited to the set $J$ (see \cite{FullPaper}).

\section{The proposed rate adaptive scheme}\label{sec:rate_adaptive_scheme}
The following communication scheme sends $B$ indices from $\{1,\ldots,M\}$ over $n$ channel uses (or equivalently sends the number $\theta \in [0,1)$ in resolution $M^{-B}$), where $M$ is fixed, and $B$ varies according to empirical channel behavior. The building block is a rateless transmission of one of $M$ codewords ($K \equiv \log(M)$ information units), which is iterated until the $n$-th symbol is reached. The codebook $C_{M \times n}$ consists of $M$ codewords of length $n$, where all $M \times n$ symbols are drawn i.i.d. $\sim Q$ and known to the sender and receiver.

In each rateless block $b=1,2,\ldots$, a new index $i = i_b \in \{1,\ldots,M\}$ is sent to the receiver. $k$ denotes the absolute time index $1 \leq k \leq n$. Block $b$ starts from index $k_b$, where $k_1=1$, and $b$ is incremented following the decoder's decision to terminate a block. After symbol $n$ is reached the transmission stops and the number of blocks sent is $B=b-1$. The transmission of each block $b$ follows the procedure described below:
\begin{enumerate}
\item The encoder sends index $i=i_b$ by sending the symbols of codeword $i$ : $x_k = C_{i,k}$, and incrementing $k$ until the decoder announces the end of the block. Note that different blocks use different symbols from the codebook.
\item The decoder announces the end of the block after symbol $m$ in the block ($m = k - k_b + 1$) if for any codeword $x_i$ :
\begin{equation}\label{eq:termination_condition}
\Remp \left( (\vr x_i)_{k_b}^k, \vr y_{k_b}^k \right) \geq \mu^*_m
\end{equation}
where $\Remp(\vr x, \vr y)$ defined in Eq.(\ref{eq:def_mu}) is used as the decoding metric and $\mu^*_m$ is a threshold defined in Eq.(\ref{eq:termination_threshold}) below.
\item When the end of block is announced one of the $i$ fulfilling Eq.(\ref{eq:termination_condition}) is determined as the index of the decoded codeword $\hat i_b$ (breaking ties arbitrarily).
\item If symbol $n$ is reached without fulfilling Eq.(\ref{eq:termination_condition}), then the last block is terminated without decoding.
\end{enumerate}

The threshold $\mu^*_m$ is defined as:
\begin{equation}\label{eq:termination_threshold}
\mu^*_m =
 \left\{ \begin{array}{ll}
\frac{ K + \log \left( \frac{n}{P_e} \right) + |\mathcal{X}||\mathcal{Y}| \log(m+1) }{m}  & \textrm{discrete}\\
\frac{ K + \log \left( \frac{2n}{P_e} \right)} {m-1} & \textrm{continuous}\end{array} \right.
\end{equation}

The scheme achieves the claims of Theorem \ref{theorem:adaptive_rate_combined} when $K$ is chosen to increase as $O(\log(n)) < K < O(n)$. The scheme uses one bit of feedback per channel use, however the same asymptotical rates are obtained if (a possibly delayed)) feedback is sent only once every $T$ symbols (for any $T>0$), therefore we can claim the theorem holds with "zero rate" feedback.

\section{Outline of the proof of the main result}\label{sec:proof_outline}
The error analysis is based on two lemmas (Lemma 1 and Lemma 4 of \cite{FullPaper}) which state that the probability of the metric $\Remp$ used in section \ref{sec:rate_adaptive_scheme} to exceed a given threshold $t$ when the $m$-length $\vr x$ is drawn i.i.d. independently of $\vr y$ (and from a Gaussian distribution in the continuous case), is approximately $\exp(-mt)$, or more accurately:
\begin{equation}
Q^m(\Remp(\vr x, \vr y) \geq t) \leq
 \left\{ \begin{array}{ll}
\exp \left( -m (t - \delta_m) \right)  & \textrm{discrete}\\
2 \exp \left( -(m-1) t \right) & \textrm{continuous}\end{array} \right.
\end{equation}
where $\delta_m = |\mathcal{X}||\mathcal{Y}|\frac{\log(m+1)}{m}$. This bound determines the pairwise error probability. Using this bound with $t=\mu^*_m$ and the union bound (over $M-1$ competing codewords and over $n$ decoding attempts), we show that the error probability is bounded below $P_e$.

The analysis of the rate is more intricate. Basically it relies on the fact that when a block is decoded, the metric $\Remp$ exceeds the threshold $\mu^* \approx \frac{K}{m}$ at the last symbol, but it lies below the threshold at the previous symbol, therefore roughly speaking, at the end of the block $\Remp \approx \mu^* \approx \frac{K}{m} = R_b$ where $R_b$ is the instantaneous transmission rate over the block. Therefore the empirical rate is attained per rateless block and a convexity argument is used in order to show that the total rate (average of $\Remp$ over blocks) is at least the empirical rate $\Remp(\vr x, \vr y)$ measured over the complete sequences.

There are several difficulties, however. Considering for example the discrete case, since the rate achieved instantaneously over a rateless block is approximately the empirical mutual information over the block, we would like to claim that the averaged rate over rateless blocks is greater or equal to the empirical mutual information over the entire transmission, which implies convexity of the empirical mutual information. However the mutual information is \emph{concave} with respect to the input distribution. Here, the input distribution is the empirical distribution over rateless blocks, whose limits are determined during transmission by the decoding rule and depend on the channel output $\vr y$. Another difficulty is that the last symbol of each block is not fully utilized: the empirical mutual information crosses the threshold at the last symbol. But whether it crosses it just barely, or crosses it by a significant extent, the rates our scheme attains remain the same. However a large increase in empirical mutual information at the last symbol increases the target rate thus increasing the gap between the target rate and the rate attained. Here, a "good" channel is bad for our purpose. Since we operate under an arbitrary channel regime, this increase is not bounded by the average information contents of a single symbol. This is especially evident in the continuous case where $\Remp$ is unbounded. A similar difficulty arises when bounding the loss from the potentially unfinished last block in the transmission: since $\vr y$ is arbitrary it can be determined so that this block has the best mutual information.

We resolve the aforementioned difficulties by proving a property we term "likely convexity" (Lemmas 5,6 in \cite{FullPaper}): given a partitioning of the symbols $1,\ldots,n$ into subsets, we show that if the number of subsets does not grow too fast, and independently of their size, there is a group of $\vr x$ sequences $J$ with vanishing probability, such that if $\vr x \not\in J$, the mutual information (in the discrete case) and the squared correlation factor (in the continuous case) are convex up to an arbitrarily small offset $\Delta$, i.e. the convex combination of mutual information (resp. $\hat\rho^2$) over the subsets, weighted by their size, exceeds the mutual information (resp. $\hat\rho^2$) measured over the entire $n$ symbols minus $\Delta$. The likely convexity is used to bound the loss from unused symbols (by bounding their number, and the mutual information or correlation factor, resp.), as well as show that the mean rate over rateless blocks meets or exceeds the overall empirical rate (mutual information or its Gaussian counterpart). Convexity of $\Remp = -\half \log(1-\hat\rho^2)$ follows from convexity of $\hat\rho^2$ by Jensen's inequality.

The likely convexity property results in the existence of the subset $J$ of bad sequences. An example for such a sequence is the sequence of $\half n$ zeros followed by $\half n$ ones (for the binary channel), in which at most one block will be sent, and thus the asymptotic rate tends to $0$, although the empirical distribution is $Ber(\half)$ ($\hat H(\vr x) = 1$) and the empirical mutual information may be $\hat I = 1$.

\begin{figure}
\center
  \includegraphics[width=8cm]{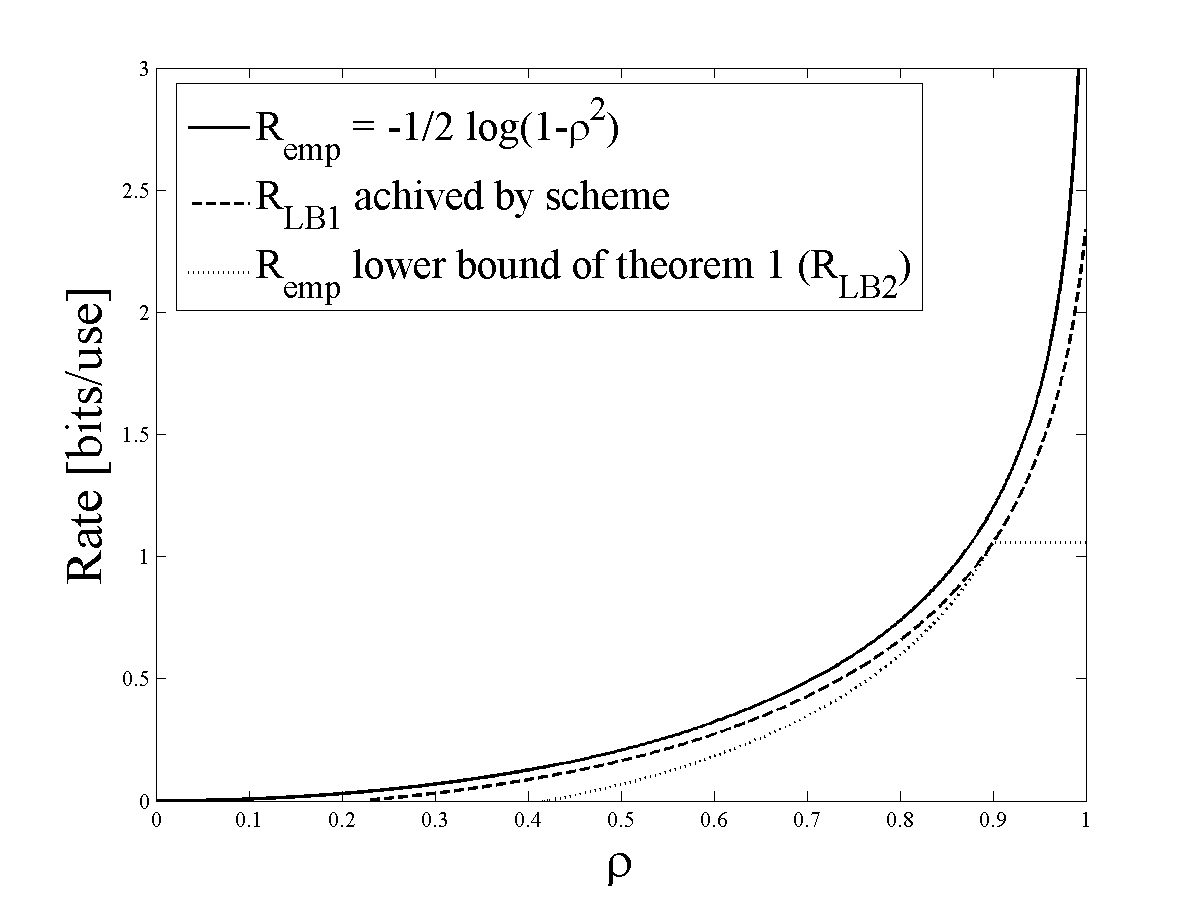}
  \caption{Illustration of $\Remp$ lower bound of theorem \ref{theorem:adaptive_rate_combined} for the continuous case ($\RLBTWO$) and the lower bound $\RLBONE$ shown in the proof in \cite{FullPaper}, as a function of
$\rho$, for $n=10^8, K=10^6, P_A=0.001, P_e=0.001$}\label{fig:rate_in_continuous_theorem}
\end{figure}

Finally, in order to make sure the error probability, the probability of $J$, and the various rate offsets inserted by the communication system and by the proof technique all tend to zero as $n \to \infty$, the information contents of each block is required to increase at a rate $O(\log(n)) < K < O(n)$. As part of the proof in \cite{FullPaper} we introduce several lemmas which seem to constitute fundamental and useful tools in analyzing individual sequences. Figure (\ref{fig:rate_in_continuous_theorem}) illustrates a lower bound for the rate achieved by the proposed scheme for finite $n$ (termed $\RLBONE$) which is calculated in \cite{FullPaper}, as well as a bound ($\RLBTWO$) satisfying the form defined in Theorem \ref{theorem:adaptive_rate_combined}.

\section{Examples}\label{sec:examples}
In this section we give some examples to illustrate the model developed in this paper. Further details appear in \cite{FullPaper}.

\subsection{Non linear channels}\label{sec:example_non_linear}
The expression $\half \log \left( \frac{1}{1-\rho^2} \right)$ determines a rate which is always achievable using a Gaussian prior, and is useful for analyzing non linear channels. As an example, transmitter noise generated by power amplifier distortions is usually modeled as an additive noise, although it is correlated with the transmitted signal, resulting in an overly optimistic model. Using the procedure described in the overview of finding the coefficient $\alpha$ such that this noise is orthogonal to the transmitted signal we can model the non linearity as an effective gain plus an additive noise. The rates computed using this model are always achievable, and thus are a practical alternative to calculating the channel capacity, and enable simplified modeling of the distortion as an additive noise.

\subsection{Channels that fail the zero order and the correlation model}\label{sec:example_failure_models}
The fact we used the zero-order empirical distribution makes the scheme less effective for channels with memory. For example for the error free channel $y_k = x_{k-1}$ the achieved rate would be 0 (with high probability). Similarly for the correlation model if $y_k = x_k^2$ then $\rho=0$. The remedy should be sought in employing higher order empirical distributions and in the continuous case in using tighter approximations of the empirical statistics (e.g. by higher order statistics).

\subsection{Application to other channel models}\label{sec:example_adversary}
As we noted in the overview, the results obtained for the arbitrary channel model constitute a convenient starting point for analyzing channel models which have a full or partial probabilistic behavior. It is clear that results regarding achievable rates in fully probabilistic, compound, arbitrarily varying and individual noise sequence models can be obtained from applying the weak law of large numbers to Theorem \ref{theorem:adaptive_rate_combined} (limited, in general, to the randomized encoders regime). For example the result of \cite{Ofer_BSC} for the binary channel $y_n=x_n \oplus e_n$ can be easily reconstructed by applying the scheme with $Q = Ber(\half)$, asymptotically approaching (or exceeding) the rate:
\begin{multline}
 \Remp = \hat I (\vr x; \vr y) = \hat H (\vr y) - \hat H (\vr y | \vr x) =
 \hat H (\vr y) - \hat H (\vr e | \vr x)
 \geq \\ \geq
 \hat H (\vr y) - \hat H (\vr e) = \hat H (\vr y) - h_b(\hat \epsilon) \arrowexpl{prob.} 1_{\mathrm{bit}} - h_b(\hat \epsilon)
\end{multline}
Since $X_k$ is i.i.d. $Ber(\half)$ so is $\vr Y_k$, and the limit follows from the law of large numbers and the continuity of $H(\cdot)$. In \cite{FullPaper} we consider the discrete channel with state sequence presented by \cite{Eswaran} where the sequence is potentially determined by an adversary knowing the past channel inputs and outputs (as opposed to a fixed sequence assumed in \cite{Eswaran}), and show by similar arguments that the same communication rates can be attained. This result is a superset of the results of \cite{Eswaran} and \cite{Ofer_EMP}, and is new, to our knowledge. Applying Theorem \ref{theorem:adaptive_rate_combined} the proof is simple: it only remains to show through a probabilistic calculation that the difference between the empirical mutual information and the target rate (the state averaged mutual information defined in \cite{Eswaran}) converges to 0 in probability.

A more anecdotic particular case is the additive Gaussian channel where by Theorem 2 and Lemma 4 of \cite{FullPaper} we obtained a very simple proof for the achievability part of this channel's capacity, using simple and geometrical considerations without the heavy machinery of AEPs or error exponents, and by employing a maximum correlation factor decoder rather than the maximum likelihood (minimum Euclidian distance) decoder.

\section{Further study}\label{sec:comments}
This work lays the foundations and introduces the new concept of individual channels together with basic achievability results. Following that, there are many open questions. To name the most outstanding ones:
\begin{itemize}
\item Extensions of the model to include time dependency
\item Definition of the empirical mutual information for continuous alphabets, and extension of the scheme to approach this empirical mutual information. Unification of the discrete and continuous cases, and extension to multiple input/output channels.
\item Analysis of the overheads and their dependence on the model complexity (asymptotical rate - overhead tradeoff)
\item Best asymptotical error rates
\item Determining and adjusting the channel input to channel behavior (e.g. by adjusting the prior), and considering alternatives to the strict constraint imposed here on the input prior
\item Outer bounds on achievable rates
\item The minimal amount of randomization required to attain the empirical mutual information
\end{itemize}
In \cite{FullPaper} we give additional details and make some initial comments about these directions.



\end{document}